\documentclass[twocolumn,showpacs,preprintnumbers,amsmath,amssymb]{revtex4}

\usepackage{graphicx}
\usepackage{bm}      

\usepackage{subfigure}
\usepackage[T1]{fontenc}
\usepackage{ae,aecompl}
\newcommand{\tstar}{t^{*}}
\newcommand{\Tstar}{T^{*}}
\newcommand{\Pstar}{P^{*}}
\newcommand{\mustar}{\mu^{*}}

\begin{document}

\title{Progression of Phase Behavior for a Sequence of Model Core-Softened Potentials}

\author{D Quigley}
\email{dq100@york.ac.uk}
\author{MIJ Probert}
\affiliation{
University of York\\
Heslington \\
York \\
United Kingdom \\
YO10 5DD 
}

\date{\today}

\begin{abstract}
A series of phase diagrams is obtained in three dimensions for a smooth pair-potential with an outer well and a repulsive inner shoulder. Condensed phase boundaries are located using free energy calculations. Liquid-vapour equilibria are obtained with multi-canonical methods. As the depth of the outer well is increased, a simple-hexagonal to close packed transition appears in the solid leading to a discontinuity in the slope of the melting curve. For deeper wells the simple hexagonal melting temperature exhibits a maximum with respect to pressure. The onset of the predicted metastable isostructural transition is also studied. 
\end{abstract}

\pacs{61.20.Ja, 61.66.Bi, 64.70.Dv, 64.70.Fx}

\maketitle


\section{Introduction}

Since the work of Stell and Hemmer \cite{StellHemmer70} it has been appreciated that a simple model pair-potential with two repulsive regions of differing strength is capable of generating a third fluid phase. Recent experimental and theoretical evidence for liquid-liquid phase transitions (LLPT) in elemental melts \cite{KatayamaMUSYF00,Morishita01a,GlosliR99,BrazhkinPV99} has revived interest in these models as a mechanism for reproduction and study of the general phenomenon. In particular, it has been suggested by \citeauthor{MishStan2} \cite{MishStan2} that so called `core-softened' potentials are capable of generating a first-order phase transition within a supercooled liquid. LLPTs of this kind are observed in many models of water \cite{PooleSES92,PooleSES93,HarringtonZPSS97,Yam02,Brov03,Brov05} and may be related to the celebrated density maximum at $4^{\circ} C$. The LLPT is considered an extension of the transition between two amorphous ice phases into the supercooled liquid regime, ending in a critical point below the thermodynamic melting temperature. 

\begin{figure} \centering
\includegraphics*[scale=0.4]{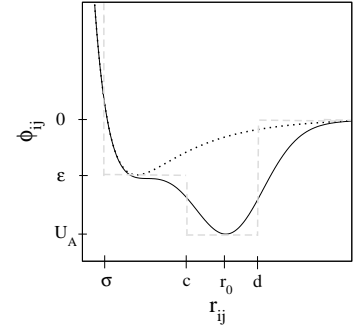}
\caption{Core softened pair-potentials. The discrete form is shown as a dashed line, along with a corresponding smooth form (solid line) constructed as in equation \ref{eq:SHpots}. The Lennard-Jones potential (dotted line) is shown for comparison.} 
\label{fig:SHpots}
\end{figure}

A continuous interpretation of a Stell-Hemmer like potential can be reproduced by the function

\begin{eqnarray}
\label{eq:SHpots}
\Phi\left(\mathbf{r}^{N}\right) &=& \sum_{j>i}^{N}
4\epsilon\left[\left(\frac{\sigma}{r_{ij}}\right)^{12} 
-\left(\frac{\sigma}{r_{ij}}\right)^{6}\right] \nonumber \\ &-&A\exp\left[-w\left(r_{ij}-r_{0}\right)^{2}\right],
\end{eqnarray}
which for appropriate choices of $A$, $r_{0}$ and $w$ closely resembles the form suggested by \citeauthor{MishStan2}. \citeauthor{ScalaSGBS00} \cite{ScalaSGBS00} have presented a clear argument that potentials of this form posses an isostructural transition between either two solid or two liquid phases of different density. Previous simulation studies have employed both the smooth form given by equation \ref{eq:SHpots} and a discrete piecewise interpretation. Both are shown in Fig. \ref{fig:SHpots}. In two dimensions the observation of a water-like density anomaly has been reported in molecular dynamics simulations of both the smooth and discrete forms \cite{Sadr-LahijanySBS98,ScalaSGBS01}. Other simulations have employed advanced Monte-Carlo methods \cite{WildingM02}. These indicate that rather than being related to a metastable LLPT, the density anomaly is a consequence of cluster formation during quasi-continuous freezing to a solid of lower density than the surrounding liquid. 

Simulations in three dimensions have been largely limited to the discrete potential, and indicate a second critical point can occur for certain parameterisations \cite{FranzeseMSBS01,BuldyrevFGMSSSS02,FranzeseMSBS02,Skibinsky04}, without a density anomaly. This metastable second critical point can lie at higher or lower temperature than the liquid-gas critical point depending upon the specific parameterisation employed.

Previous simulations of the \emph{smooth} potential in three dimension have been restricted to two limited studies \cite{MMay03,Netz04}. In a recent rapid communication \cite{QuigleyP05}, we mapped the low pressure portion of the three-dimensional phase diagram for $A=\epsilon$. A simple hexagonal (sh) solid phase, identified using the meta-dynamics method of \citeauthor{ParRarRevisit} \cite{ParRarRevisit}, was shown to remain thermodynamically stable in this regime. In this paper we report on calculations for a range of $A$ values, studying the emergence of the sh phase. The sh to close packed transition at high pressures is studied. In addition we follow the behaviour of the melting and liquid-vapour critical temperatures as a function of the outer well depth. The possibility of isostructural phase transitions is also investigated.

Measured quantities are presented here in the usual dimensionless reduced units. Energies are quoted as multiples of the inner-well depth $\epsilon$, lengths as multiples of $\sigma$. Reduced temperature $T^{*}$ is calculated as $k_{B}T/\epsilon$, with pressure $P^{*}=P\sigma^{3}/\epsilon$. Time is measured in units $t^{*}=(m/\epsilon)^{1/2}\sigma$ where $m$ is the atomic mass. For all studies reported here the parameters $r_{0}=1.44\sigma$ and $w=41.22\sigma^{-2}$ are used in equation \ref{eq:SHpots}. The pair-potential is truncated at $2.5\sigma$ and \emph{force}-shifted such that no discontinuities occur in the dynamics.

The remainder of this paper is organised as follows. In section \ref{sec:zeroT} we explore the zero temperature phase behaviour as a function of the outer well depth $A$ and choose values for study at finite temperature. In section \ref{sec:methods} we describe the methods employed for our finite temperature studies. Results are presented in section \ref{sec:results}. Finally, we summarise our findings.

\section{Zero Temperature Behaviour}
\label{sec:zeroT}

\begin{figure}
\centering
\includegraphics*[scale=0.38]{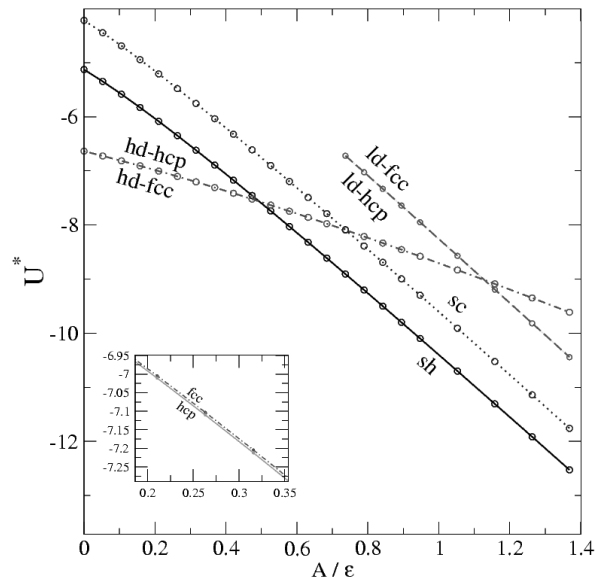}
\caption{Optimised energy at zero temperature and pressure for various structures as a function of the outer well depth parameter. The hcp and fcc structures are near-degenerate as indicated in the inset.}
\label{fig:optva1}
\end{figure}

Much insight into the phase behaviour of a model substance can be extracted from zero temperature enthalpy and volume characteristics. The zero temperature energy is obtained by performing a conjugate-gradient (CG) minimisation of the potential energy  $U$ with respect to both the cell vectors and the fractional atomic coordinates. Our identification of candidate structures was reported in Ref. \cite{QuigleyP05}. The energy under zero pressure for energetically relevant structures is plotted as a function of the outer well depth $A$ in Fig. \ref{fig:optva1}. In order of decreasing density, these are the high density face centred and hexagonal close packed structures (hd-fcc and hd-hcp) associated with the unmodified Lennard-Jones potential, simple hexagonal (sh), simple cubic (sc) and a second set of lower density close packed structures (ld-fcc and ld-hcp). The final two structures are stabilised by the outer well only for $A>0.7\epsilon$.

The energetic favourability of open structures on increasing $A$ can be understood in terms of neighbour distances. The sh primitive cell is defined by $a=b=1.059\sigma$ and  $c=1.016\sigma$ with  $\alpha=\beta=90^{\circ}$,$\gamma=120^{\circ}$. The nearest and second nearest neighbours lie at a distance $c$ and $a$ respectively, close to the position of the Lennard-Jones minimum, while the third nearest lie at $\sqrt{\left(a^{2}+c^{2}\right)}$ which is very close to $r_{0}$. Both energy minima in the pair potential are therefore utilised. In the sc structure, nearest neighbours lie at approximately $1.12\sigma$. The favourability of sc over the high density close packed structures stems from second nearest neighbours which lie at $\sqrt{2}$ times this distance, which is close to $r_{0}$. Each atom has six neighbours close to the Lennard-Jones minimum plus six close to $r_{0}$. This compares to eight and twelve in the sh case which is therefore lower in energy.

Both pairs of close-packed structures decrease in energy on increasing $A$. The lower density structures contain more neighbours close to $r_{0}$ and hence the decrease is faster in these cases.

\section{Methods}
\label{sec:methods}
\subsection{Liquid-Vapour Equilibria}
\label{sec:methods-liqgas}

Liquid-vapour equilibria are traced in the direction of decreasing temperature from near the critical point. An initial estimate of this critical point is obtained using molecular dynamics simulations in the canonical ensemble. Fluid isotherms are traced in the pressure-density plane at temperatures increasing in steps of $\Delta T^{*} = 0.08$. The temperature of highest isotherm exhibiting the classic hysteresis associated with the liquid-vapour transition is taken as an estimate of the critical temperature. Typically between 4 and 8 isotherms are required for this process. Each isotherm is computed from 30 simulations over the density range $\rho^{*}=0.1$ to $0.8$. Simulation time at each density is $t^{*}=400$ in both equilibration and production periods.

Tracing of the liquid-vapour coexistence curve proceeds using Monte-Carlo simulations in the grand-canonical ensemble (GCE). These combine multi canonical sampling \cite{multican} with re-weighting of particle number histograms \cite{FerrenbergS88}. Our method closely follows that described in detail by \citeauthor{WildingTutorial} \cite{WildingTutorial}. A starting point on the liquid-vapour coexistence curve is obtained from the above estimate of critical temperature by fine tuning of the chemical potential $\mu^{*}$ until a bimodal distribution is measured in the particle number $N$, with equal area in each peak. This is then re-weighted to a lower temperature, providing a biasing function for multi-canonical sampling. The process repeats, proceeding down the coexistence curve in steps of $\Delta \Tstar=-0.008$. At each step a multi-canonical GCE simulation of $500$ $000$ Monte-Carlo cycles is sampled at equilibrium to produce the particle number histogram. A cubic simulation cell of side $L=7.13\sigma$ is employed in all cases. Typically twenty steps are employed, tracing the curve over a temperature range of approximately $\Delta \Tstar=0.15$. The resulting data is used to estimate the parameters of the liquid-gas critical point.

At lower temperatures the liquid-vapour curve is traced by numerical integration of the Clausius-Clapeyron equation (see below).

\subsection{Solid-Solid Transitions}

These are located via free energy calculation of both phases along an appropriately chosen isotherm. Free energies are calculated using the Einstein crystal method of Frenkel and Ladd \cite{FrenkelLadd}. Here the free energy derivative is integrated by sampling typically $20-50$ points along the path connecting the core-softened solid to the reference harmonic crystal, using canonical ensemble Langevin dynamics simulations. Each sample employs a total simulation time $t^{*}=3.6$ at equilibrium. The appropriate cell in which to perform this thermodynamic integration is first identified by employing constant pressure Langevin dynamics simulations \cite{QuigleyP04} of typical duration $\tstar=100$ after equilibration. A fully flexible simulation cell is used in all cases. System sizes of $N=256$ and $392$ use used for fcc, sh structures respectively.

Two sources of system size dependence arise in these calculations. The first arises in the thermodynamic integration itself. This has been largely compensated for with repeat integrations at larger $N$. The free energy is then extrapolated to the thermodynamic limit via the method of \citeauthor{FrenkelLaddFS} \cite{FrenkelLaddFS}. The second arises from finite-size effects in the identified density at a given pressure. A selection of repeat free energy calculations at densities obtained from larger system sizes have shown this to be negligible. In both cases the additional system sizes used are $N=500$ and $864$ for fcc, $N=640$ and $864$ for sh. 

Free energy calculations of hcp and sc structures are also reported below, but are not used in locating phase boundaries. No finite size analysis has been performed in these cases. System sizes used are $N=384$ for hcp, and $N=343$ for sc.

\subsection{Melting Curves}
\label{sec:methodmelt}

Melting curves are identified by two independent methods in this work. The first is based on direct simulation of phase coexistence in the $NPT$ and $NPH$ ensembles \cite{MorrisMelt}. The $NPT$ ensemble method is used here to study melting of fcc structures. A cell is prepared with $500$ atoms in each phase and simulated at a specified temperature and pressure for $\tstar=300$. During this time the system transforms to a pure solid or liquid state. A series of these simulations allow the melting temperature to be accurately bracketed.

In contrast, the $NPH$ method requires a single simulation to locate the melting temperature at a specified pressure \cite{Wang05}. Provided the initial enthalpy of the system lies close to that at the melting temperature, fractions of the system will melt or solidify, shifting portions of the conserved enthalpy between the two phases. The thermodynamic melting temperature is obtained at equilibrium. We note that $NPH$ simulations employing an Andersen-Hoover \cite{Andersen80,Hoover86} or (as in this work ) a Martyna-Tobias-Klein \cite{MTKCorrection} barostat, the enthalpy is \emph{not} exactly conserved, but is constant to within fluctuations in the fictitious kinetic energy associated with the cell dynamics. This pseudo-NPH approach is therefore only useful in cases where this fluctuation can be kept within the latent heat per atom associated with the melting transition. This criteria has only achievable for melting of the simple hexagonal structure. In this case a simulation cell with $332$ atoms in each phase is prepared and simulated as described in Ref. \cite{QuigleyP05}.

Melting temperatures for each potential are also computed from solid and liquid free energies along an appropriately chosen isobar. Solid free energies are computed as above with finite-size corrections applied in all cases, up to the maximum temperature of mechanical stability. Liquid free energies are computed via thermodynamic integration from a reference point of known chemical potential. This is obtained as follows.

A point is chosen within the liquid region of the $T-\mu$ plane. The choice is aided by the data obtained during the multi-canonical sampling, and by the solid free energy calculations which yield the chemical potential of the superheated solid. At the chosen reference point a further GCE Monte-Carlo simulation is conducted to obtain density and pressure information. The Helmholtz free energy at the reference point can then be calculated to within the statistical uncertainty inherent in this simulation. These GCE simulations employ $500$ $000$ MC cycles in a cubic simulation cell of side $7.13\sigma$.

Liquid free energies along the isobar of interest are obtained using thermodynamic integration to the density (identified using constant pressure Langevin dynamics simulations as for the solid) and temperature of interest. Each integration samples the relevant free energy derivative at 10 points along an isotherm and isochore using canonical ensemble Langevin dynamics simulations of duration $t^{*}=50$. A system size of $N=350$ is used.

Errors in the thermodynamic integration have been estimated by evaluating the change in free energy around various closed loops in the $T-\rho$ plane. The system size dependence has been estimated by repeating a selection of integrations with $N=600$, and by employing larger simulation cells in GCE simulations. The dominant source of error is identified as statistical uncertainty in the liquid reference point. This has been controlled by employing suitably long GCE simulations.

\subsection{Clausius-Clapeyron Integration}

The free energy calculations described above locate a single point on a phase boundary. To trace the remainder of a phase coexistence curve, the Gibbs-Duhem integration method \cite{Kofke} is employed. Beginning from the identified single point, the Clausius-Clapeyron equation is evaluated over two (one for each phase) constant pressure Langevin dynamics simulations. This is then integrated with the fourth-order Runge-Kutta method to trace the phase boundary in the $P-T$ plane. Simulations at each point are typically of duration $\tstar=100$ to $150$. System sizes are $N=500$ for fcc phases and $N=640$ for sh phases. In the case of melting curves the number of atoms in the liquid matches that in the solid. 

For tracing liquid-vapour coexistence curves a system size of $N=500$ is used for both phases. An initial point for the series is taken from a fit to temperature-pressure data unfolded from the multi-canonical GCE simulations.

Integration error is estimated by reversing each series, integrating back toward the initial point. In all cases this error is much smaller than the uncertainty of the initial point.

\section{Results}
\label{sec:results}

\subsection{Phase Behaviour for $A=\epsilon/4$}

At $A=\epsilon/4$ the outer Gaussian well represents a small perturbation on the Lennard-Jones potential. No significant change in phase behaviour is therefore expected. 
\begin{figure}
\centering
\subfigure[]{
\includegraphics*[scale=0.52]{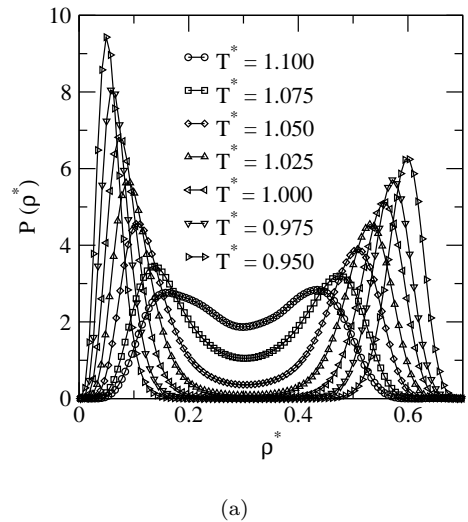}
\label{fig:sh1liqgas}
}
\subfigure[]{
\includegraphics*[scale=0.5]{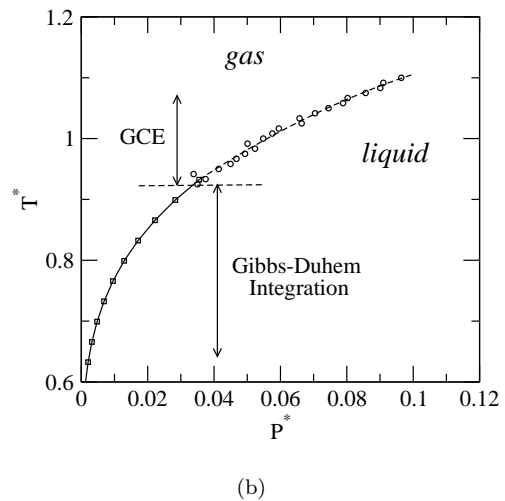}
\label{fig:sh1join}
}
\caption{Liquid-gas transition in the $A=\epsilon/4$ potential. Sample histograms obtained during multi-canonical GCE simulations are shown in (a). At lower temperatures the transition is traced using Gibbs-Duhem integration. The join between the two methods in the $P-T$ plane is shown in (b).}
\end{figure}
Following the procedure in section \ref{sec:methods-liqgas}, an initial bimodal particle number histogram was obtained at $T^{*}=1.100$ with a chemical potential of $\mu^{*}=-9.267$. The liquid-vapour curve was traced to a temperature of $T^{*}=0.925$ in 21 steps. Example histograms are shown in Fig. \ref{fig:sh1liqgas}. These yield information on the density of the two phases, allowing a fit to the scaling law 
\begin{equation}
\rho^{*}_{l}-\rho^{*}_{g} \propto (\Tstar-\Tstar_{c})^{\beta}
\end{equation}
and to the law of rectilinear diameters,
\begin{equation}
\frac{\rho^{*}_{l}+\rho^{*}_{g}}{2} = \rho^{*}_{c}+A(\Tstar-\Tstar_{c}).
\end{equation}
This process identifies the critical point at $T_{c}^{*}=1.108\pm 0.003$, $\rho^{*}_{c}=0.302\pm0.003$. Note that the finite-size dependence of these parameters has not been investigated.

Below $T^{*}=0.925$ the continuation of the liquid-vapour coexistence curve has been traced with Gibbs-Duhem integration. The continuation is shown in Fig. \ref{fig:sh1join}.

Free energy calculations have been performed at temperatures in the range $T^{*}=0.083$ to $0.500$ in steps of $0.042$ along the zero pressure isobar. At no temperature in this range were the sc and sh structures mechanically stable. As the fcc structure is of the highest density, we can conclude that as with the Lennard-Jones system, the solid remains close packed over the entire positive pressure range.
\begin{figure}
\centering
\includegraphics*[scale=0.5]{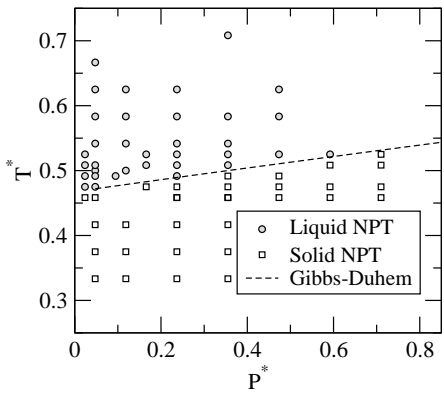}
\caption[Melting curve for the $A=\epsilon/4$ potential.]{Melting curve for the potential with $A=\epsilon/4$ calculated using the two-phase NPT coexistence method described in section \ref{sec:methodmelt}. The dashed line is the melting curve computed from free energy calculations and traced with Gibbs-Duhem integration.}
\label{fig:sh1meltnpt}
\end{figure}

Results of two-phase $NPT$ melting calculations close to the melting line are shown in Fig. \ref{fig:sh1meltnpt}. 80 simulations were conducted in total. Each sampled point is visually identified as either solid or liquid. Simulations which could not be identified as a pure phase within the simulation time (i.e. those close to the melting line) are not plotted. 

The melting temperature at a pressure of $P^{*}=0.047$ has been calculated accurately by free energy calculation. For the solid phase, calculations were performed a temperatures up to $T^{*}=0.500$ in steps of $0.042$. A reference point for liquid free energy calculations was taken at $T^{*}=1.000$ and total chemical potential $\mu^{*}=-10.526$. GCE Monte-Carlo simulation allows the Helmholtz free energy at this point to be identified as $f^{*}=-11.080 \pm 0.003$ per atom. Six free energies along the $P^{*}=0.047$ isobar have been computed by thermodynamic integration from this reference point, at temperatures of $T^{*}=0.375$ to $0.583$. Interpolation to the intersection reveals a melting temperature of $T^{*}=0.471$. The total error on this melting temperature is estimated at less than $\pm 0.008$, dominated by statistical uncertainty in the pressure and density of the liquid reference point.

\begin{figure}
\centering
\includegraphics*[scale=0.55]{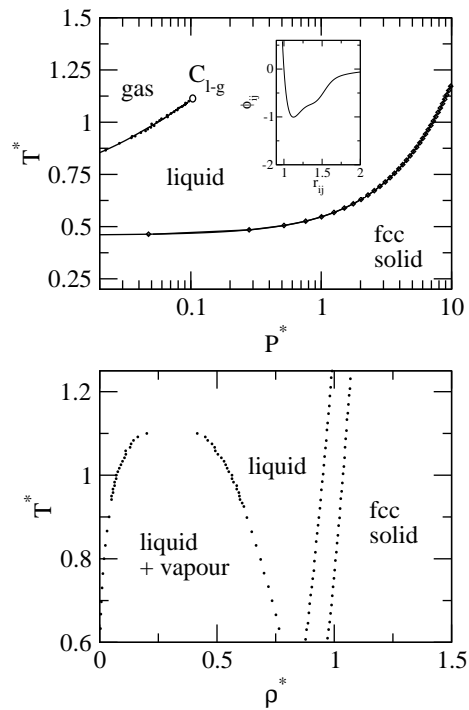}
\caption[Phase diagram of the $A=\epsilon/4$ potential.]{Phase diagram of the $A=\epsilon/4$ potential (inset). The temperature-pressure and density-temperature projections are shown. Both forward and reverse Gibbs-Duhem series are plotted. The two are indistinguishable on this scale. The low pressure region and triple point have not been studied in detail. Other than shifts in the melting and critical temperatures, no interesting phase behaviour over the Lennard-Jones case is observed.}
\label{fig:sh1phasedia}
\end{figure}

This melting point has been used to begin Gibbs-Duhem integration in the direction of increasing temperature. The resulting series is shown in Fig. \ref{fig:sh1meltnpt} and in Fig. \ref{fig:sh1phasedia} along with the liquid-vapour coexistence line.

\subsection{Phase Behaviour for $A=\epsilon/2$}

For this value of $A$ we are interested in determining if the sh structure remains thermodynamically stable at appreciable temperature, and the location of the possible sh-fcc transition.

At a temperature of $T^{*}=1.279$ a bimodal histogram on the liquid-vapour coexistence curve was obtained at $\mu^{*}=-10.782$. Subsequent histograms obtained along the transition identify the critical point at $T_{c}^{*}=1.293 \pm 0.006$, $\rho^{*}_{c}=0.284\pm0.006$.  Below temperatures of $\Tstar=1.142$ the liquid-vapour curve was traced with Gibbs-Duhem integration. 

\begin{figure}
\centering
\includegraphics*[scale=0.55]{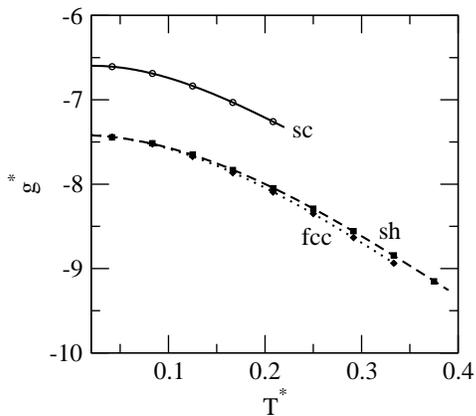}
\caption{Gibbs free energy per atom as a function of temperature along the zero pressure isobar for the $A=\epsilon/2$ potential. Three structures are shown. The sc structure is mechanically unstable beyond $T^{*}=0.25$. The fcc and sh structures are indistinguishable over much of the temperature range.}
\label{fig:sh2ti}
\end{figure}

The resulting free energies are shown in Fig. \ref{fig:sh2ti}. These are uncorrected for finite-size effects. The sc structure is now mechanically stable at finite temperature, but not beyond temperatures of $T^{*}=0.25$. Although at zero temperature the energy of the sh structure is slightly less than that of the fcc, at no finite temperature realisable in a simulation is this the case. As temperature increases the separation between the fcc and sh structures increases. There must therefore be a transition between these two structures. The free energy difference is however too low to resolve with the methods employed here.

\begin{figure}
\centering
\includegraphics*[scale=0.55]{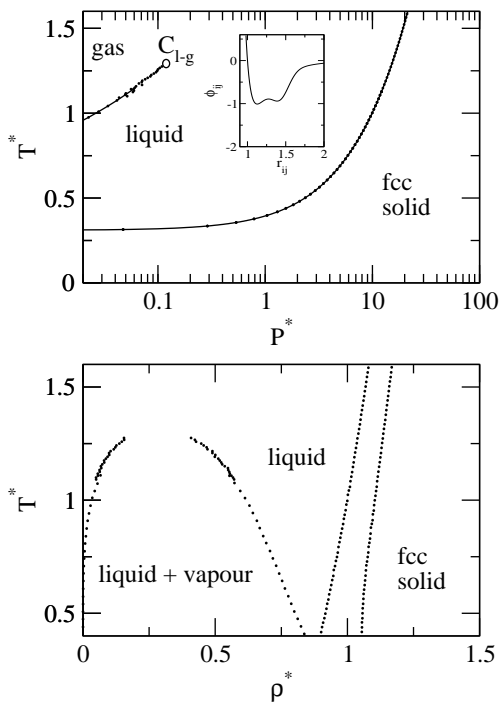}
\caption{Phase diagram of the $A=\epsilon/2$ potential (inset). The temperature-pressure projection and density-temperature projection are shown. This phase diagram represents a further decrease in melting temperature, and increase in critical temperature over the Lennard-Jones case. No new phase behaviour is observed.}
\label{fig:sh2phasedia}
\end{figure}

A total of 63 two-phase NPT simulations locate the melting temperature between $T^{*}=0.3$ and $T^{*}=0.4$ at pressures $P^{*}<0.7$. An accurate point on the melting curve was sought along the $P^{*}=0.047$ isobar. For the solid phase, free energy calculations were performed for temperatures in the range $0.208$ to $0.5$ in steps of $0.042$. For liquid free energy calculations, a reference point at $T^{*}=1.167$ with $\mu^{*}=-9.474$ was taken with a  Helmholtz free energy per atom of $f^{*}=-12.75 \pm 0.01$. Thermodynamic integration to temperatures between $0.292$ to $0.500$ in steps of $0.042$ locates the melting temperature at $T^{*}=0.315 \pm 0.01$. Again the uncertainty is dominated by statistical error in the liquid reference free energy. Gibbs-Duhem integration initiated from this point proceeded in the direction of positive temperature, producing the phase diagram presented in Fig. \ref{fig:sh2phasedia}.

\subsection{Phase Behaviour for $A=0.55\epsilon$}

It is clear that the choice of $A=\epsilon/2$ has not captured the interesting phase behaviour expected in this region, specifically the transition from sh to fcc structure. Increasing $A$ to $0.55\epsilon$ widens the energy difference between these two structures and should therefore manifest the transition at higher temperature and pressure. A study of the condensed phase behaviour of this model therefore seems appropriate.

To locate the pressure of the sh-fcc transition at zero temperature, the optimised enthalpy as a function of pressure was plotted for both structures. This enthalpy was obtained by CG minimisation with respect to both atomic positions and cell vectors. The two enthalpy curves intersect at approximately $P^{*}=8.9$. At this pressure the sh structure is of lower density. By simple consideration of the Clausius-Clapeyron equation these results require that the transition at higher temperature occurs at lower pressure. This provides a range over which the finite temperature transition can be sought with free energy calculations.
\begin{figure}
\centering
\includegraphics*[scale=0.55]{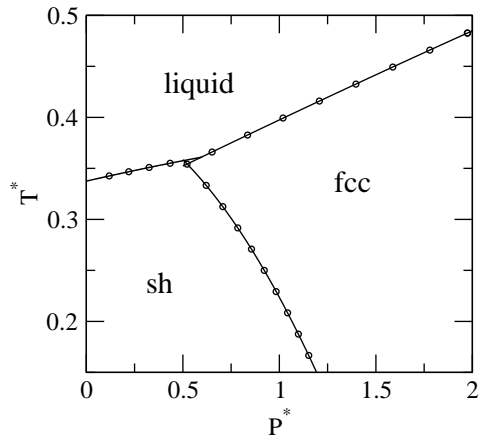}
\caption[SH-FCC-liquid triple point in the $A=0.55\epsilon$ potential.]{Phase diagram for the potential with $A=0.55\epsilon$ in the region of the sh-fcc-liquid triple point. The three Gibbs-Duhem series form a triangle at their intersection indicating the area in which the triple point is located.}
\label{fig:sh2aintersect}
\end{figure}
A finite temperature point on the sh-fcc phase boundary was sought along the $T^{*}=0.167$ isotherm. It should be stressed that density responses to pressure and temperature in the sh structure occur anisotropically. Expansion along each crystal direction must be considered separately when constructing a cell for thermodynamic integration. Free energies were calculated for pressure in the range $\Pstar=0.9$ to $1.3$. Extrapolation between sampled points yields a transition at $P^{*}=1.15\pm0.01$. This error is estimated using the deviation from ideal finite-size scaling of the leading term in $1/N$ when computing finite-size corrections.

Melting of the sh structure has been studied along the  $P^{*}=0.118$ isobar, conducting free energy calculations close to the $A=\epsilon/2$ melting line. A reference point for computing liquid free energies was taken at $T^{*}=1.171$ with chemical potential $\mu^{*}=-9.474$ and Helmholtz free energy $f^{*}=-12.95 \pm 0.01$ per atom. The intersection of the sh-solid and liquid free energies provides a melting temperature of $T^{*}=0.343\pm 0.008$. By a similar process, the fcc melting temperature at $P^{*}=2.373$ was determined as $T^{*}=0.516 \pm 0.008$.

These two melting points, plus the identified sh-fcc transition pressure at $T^{*}=0.167$ provide starting points from which to begin Gibbs-Duhem integration. The three Gibbs-Duhem series are shown in Fig. \ref{fig:sh2aintersect}. The sh-fcc meets the intersection of the two melting lines. The same triple point is located by the intersection of any two series, and is independently confirmed by the third to within a small error.  The location of the triple point is hence $\Tstar_{tp}=0.358\pm0.002$, $\Pstar_{tp}=0.54\pm0.02$.

The Gibbs-Duhem information also confirms that both melting points measured are thermodynamically stable, information which was not available from the above free energy calculations alone.

\subsection{Phase Behaviour for $A=\epsilon$}
\label{sec:aeqeps}
Here we expect the sh structure to dominate at low pressure. The transition to fcc is expected to occur at significantly higher pressures than the $A=0.55\epsilon$ case. 

A suitable bimodal number density histogram from which to begin the histogram re-weighting/multi-canonical sampling procedure was identified at $T^{*}=1.671$ when using a chemical potential of $\mu^{*}$=-14.137. Density data produced from stepping along the coexistence curve identifies the critical temperature as $\Tstar_{c} = 1.680 \pm 0.004$ when extrapolated to zero density difference. This leads to a critical density of $\rho^{*}_{c}=0.2716\pm0.0005$. Below temperatures of $\Tstar=1.450$ the coexistence curve was traced with Gibbs-Duhem integration.

\begin{figure}
\centering
\includegraphics*[scale=0.47]{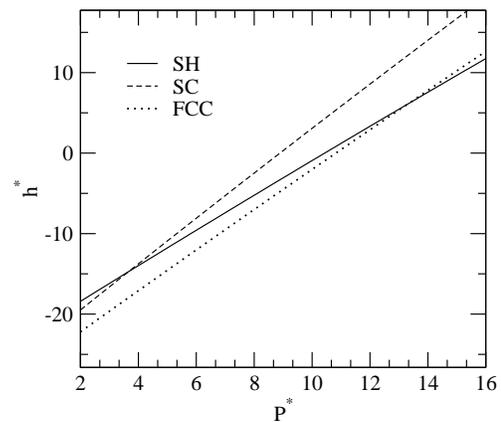}
\caption{Optimised enthalpy per atom for fcc, sh and sc structures as a function of pressure at zero temperature in the $A=\epsilon$ case. Both sc-fcc (metastable) and sh-fcc phase transitions are indicated.}
\label{fig:sh3ti2}
\end{figure}

Zero pressure free energy calculations for the relevant solid structures have been reported in Ref. \cite{QuigleyP05} and will not be repeated here. These confirm that the sh structure is thermodynamically stable at low pressure. Zero temperature phase transitions have been located by CG enthalpy optimisation of the relevant structures at various pressures. Plots of enthalpy against pressure for the sc, sh and fcc structures are shown in Fig. \ref{fig:sh3ti2}. The metastable sc-fcc transition (which is metastable) occurs at a pressure of $\Pstar=3.63$, with the sh-fcc transition at $\Pstar=13.31$.
The sh-fcc transition at finite temperature was located by performing six Einstein crystal calculations for each phase in the pressure range $\Pstar=10$ to $17$ at a temperature of $\Tstar=0.292$. The metastable sc-fcc transition has not been explored at finite temperature.

As noted in section \ref{sec:methods} the NPH two-phase method is useful in locating the sh melting temperature. These simulations in the $A=\epsilon$ case have been described in Ref. \cite{QuigleyP05}, leading to a melting temperature of $T^{*}=0.609 \pm 0.002$ at $\Pstar=0.237$. To confirm this with free energy calculation, a liquid reference point at $T^{*}=1.500$ with $\mu^{*}=-12.368$ was taken. Grand canonical Monte-Carlo simulation reveals $\left<N\right>=366.1 \pm 0.5$, and $\left<P^{*}\right>=3.59 \pm 0.01$ under these conditions. The Helmholtz free energy of the reference point is hence computed as $f^{*}=-16.99 \pm 0.01$ per atom.

Based on the information provided by the two-phase simulations, the sh melting temperature was sought along the $\Pstar=0.237$ isobar. Free energies for both phases were computed at points between $\Tstar=0.471$ and $0.671$ in steps of $0.041$. Interpolation to the intersection provides a melting temperature of $\Tstar=0.614 \pm 0.009$ which is in agreement with the two-phase NPH  result. The fcc melting  has been  studied at a pressure of $\Pstar=24.86$ by a similar procedure, locating the melting temperature at $\Tstar=1.851 \pm 0.009$.
\begin{figure}
\centering
\includegraphics*[scale=0.55]{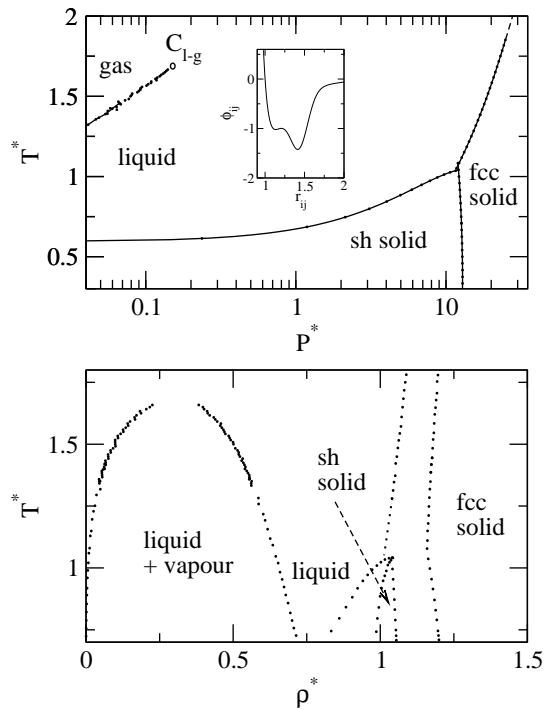}
\caption[Phase diagram of the $A=\epsilon$ potential.]{Phase diagram of the $A=\epsilon$ potential (inset) in both the pressure-temperature and temperature-density planes. The sh phase now dominates at low pressure. A maximum in the sh melting curve is marginally preempted by the fcc phase. A further increase in critical temperature with $A$ is observed.}
\label{fig:sh3fullphase}
\end{figure}

As with the $A=0.55\epsilon$ potential, the three pairs of free energy calculations have been used as starting points for Gibbs-Duhem integration. Any two of the resulting three series locate the same triple point to within the error of the initial free energy calculations. The triple point is located at $\Tstar_{tp}=1.04\pm0.01$, $\Pstar_{tp}=11.7\pm0.2$. The sh melting curve can be traced deep into the fcc region with a large range of metastability. Within this range, a maximum in the melting curve appears at slightly higher pressures than the sh-fcc triple point. 

The full phase diagram for this potential is plotted in Fig. \ref{fig:sh3fullphase}.

\subsection{Phase Behaviour for $A=3\epsilon/2$}
\label{sec:sh4}

In this potential the sh melting curve may exhibit a maximum in the stable regime. In addition, an isostructural fcc-fcc transition is expected at positive pressure.

\begin{figure}
\subfigure[]{
\includegraphics*[scale=0.48]{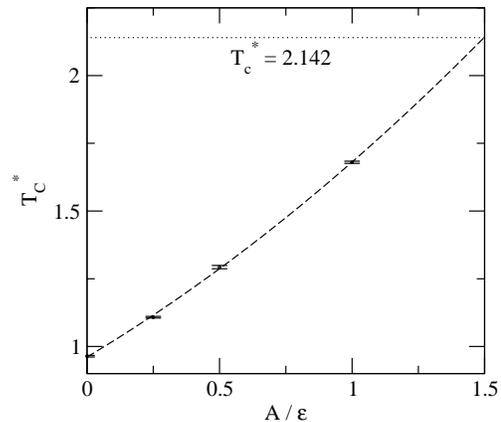}
}
\subfigure[]{
\includegraphics*[scale=0.38]{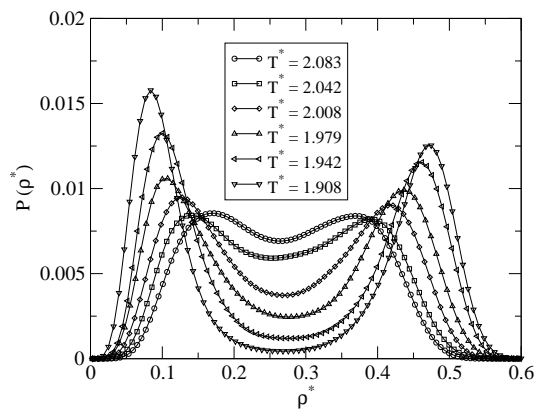}
}
\caption{Critical behaviour for the $A=3\epsilon /2$ potential. The critical temperatures identified for smaller values of $A$ are shown above. A quadratic extrapolation to the current $A$ is shown. Histograms resulting from the subsequent re-weighting and multi-canonical sampling procedure are shown below.}
\label{fig:sh4crit}
\end{figure}

Rather than employing the expensive process of tracing fluid isotherms, a starting point for the multi-canonical sampling procedure was obtained by extrapolation from previous $A$ values. Fig. \ref{fig:sh4crit} shows a plot of critical temperature against the outer Gaussian well depth. A quadratic fit to the previously identified critical temperatures yields $\Tstar_{c}=2.142$ when extrapolated to $A=1.5\epsilon$. Based on this estimate, a suitable bimodal histogram from which to begin tracing the liquid vapour coexistence curve was identified at $\Tstar=2.083$ with a chemical potential of $\mustar=-17.705$. Extrapolation of the resulting multi-canonical density data identifies the critical point at $\Tstar_{c} = 2.12 \pm 0.04$, $\rho^{*}_{c}=0.265\pm0.002$.

\begin{figure}
\includegraphics*[scale=0.6]{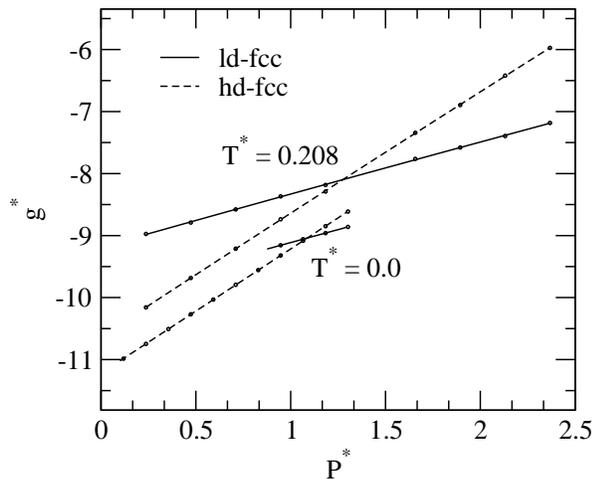}
\caption{Metastable isostructural transition in the $A=3\epsilon/2$ potential. The enthalpy per atom at zero temperature is plotted along with the results of free energy calculations along the $\Tstar=0.21$ isotherm.}
\label{fig:sh4iso}
\end{figure}

The isostructural transition was first located at zero temperature using CG enthalpy minimisation, optimising both phases at pressures between $\Pstar=0.237$ and $\Pstar=2.37$ in steps of $0.12$. The resulting enthalpy per atom for both phases is shown in Fig. \ref{fig:sh4iso}. The intersection reveals a transition pressure of $\Pstar=1.09$. Note that above $\Pstar=1.4$ the lower density phase collapses to the higher density structure during optimisation, indicating mechanical instability. 

To locate the transition at finite temperature, free energy calculations were performed for both phases along the $\Tstar=0.208$ isotherm. The chemical potentials derived from these free energies are plotted in Fig. \ref{fig:sh4iso}, locating the transition pressure for this temperature as $\Pstar=1.27$. As with sh-fcc transitions the error on this value is small, being approximately $0.01$. Both free energy series are corrected for finite-size effects.

The isostructural transition predicted by \citeauthor{ScalaSGBS00} \cite{ScalaSGBS00} has now been identified as fcc-fcc at low temperature. To locate the extent of the fcc-fcc transition, Gibbs-Duhem integration has been employed in the direction of increasing temperature.

The resulting series indicates an increase in transition pressure with increasing temperature. However, after just six steps the lower density fcc phase becomes mechanically unstable. At this point a large density difference between the two phases still exists. The isostructural transition does \emph{not} end in a critical point, but terminates at the low density fcc spinodal line.


The sh-fcc transition was located by the same procedure as in section \ref{sec:aeqeps}. The zero temperature transition is located by CG enthalpy minimisation at $\Pstar=25.27$. The transition along the $\Tstar=0.21$ isotherm was then determined by computing the free energy of both phases at ten pressures between $\Pstar=23.67$ and $\Pstar=26.04$. The transition is located at $\Pstar=25.17$. The estimated error in this pressure is of order $0.01$.

The sh and fcc melting temperature were sought along the $\Pstar=0.237$ and $23.67$ isobars respectively.
No two-phase simulations were performed in this case. A suitable range over which to perform free energy calculations was estimated from trends in earlier data. A reference point at $T^{*}=1.500$ with $\mu^{*}=-12.368$ was used in computing liquid free energies along both isobars. Grand canonical Monte-Carlo simulation provides $\left<N\right>=415.6 \pm 0.5$, and $\left<P^{*}\right>=4.52 \pm 0.01$, leading to a Helmholtz free energy of $f^{*}=-20.79 \pm 0.04$ per atom. Free energy plots along each isobar locate the sh melting temperature as $\Tstar=0.80$, with the fcc melting at $\Tstar=1.45$. In both cases the uncertainty is approximately $0.01$.
\begin{figure}
\centering
\includegraphics*[scale=0.55]{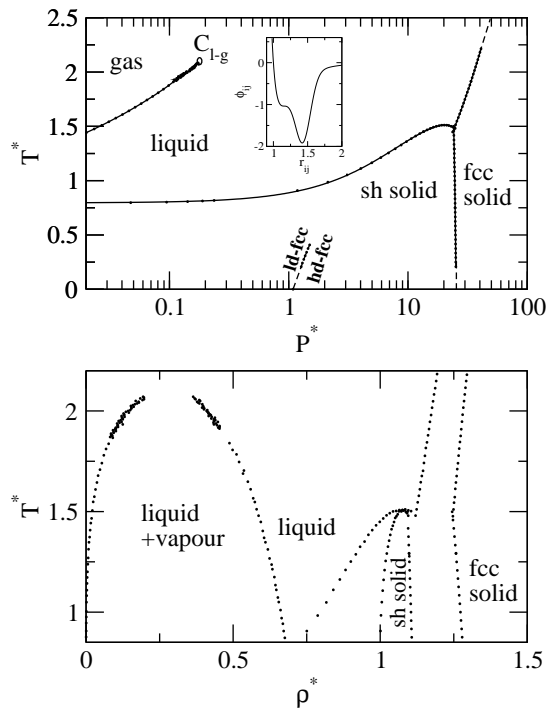}
\caption{Phase diagram of the $A=3\epsilon/2$ potential (inset). Pressure-temperature and temperature-density projections are shown. The maximum in the sh melting curve is clearly visible. The dashed line indicates the metastable isostructural transition. An increase in both melting and critical temperatures is observed over the $A=\epsilon$ case.}
\label{fig:sh4fullphase}
\end{figure}
Gibbs-Duhem series for the sh-fcc, sh-liquid and fcc-liquid transitions have been computed to trace the remainder of the phase diagram. As can be seen in Fig. \ref{fig:sh4fullphase}, the sh melting curve maximum now lies within the stable regime. The fcc melting curve intersects at slightly higher temperature. There is hence a small range visible in both the temperature-pressure and temperature-density projections for which melting is reentrant. The metastable fcc-fcc transition is also plotted in Fig. \ref{fig:sh4fullphase}. The termination of this lies well below the melting temperature.

\section{Liquid Anomalies}

In mapping the above phase diagrams, no evidence of a thermodynamically stable liquid-liquid phase transition has emerged. In particular data from Gibbs-Duhem integration reveal that density is continuous along the liquid side of all melting and vaporisation curves. In addition no third fluid peak has emerged during multi-canonical sampling. A stable LLPT must meet either the melting or vaporisation curves at a triple point and can hence be ruled out.

\begin{figure}
\centering
\includegraphics*[scale=0.47]{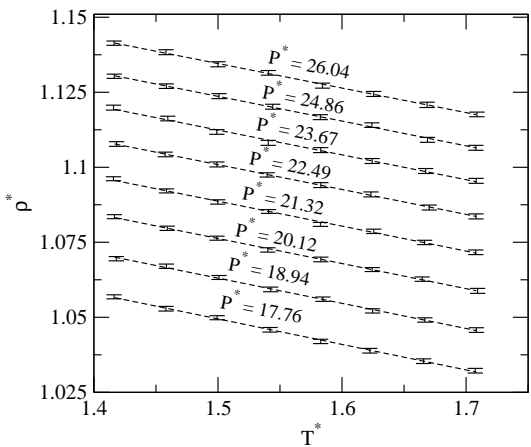}
\caption{Liquid density close to the sh melting line for $A=3\epsilon/2$. Density against temperature along several isobars is shown with accompanying error bars.}
\label{fig:denmax}
\end{figure}

The possibility of liquid anomalies has been investigated with a considerable number of constant pressure Langevin dynamics simulations. Particular attention has been focused on the $A=3\epsilon/2$ liquid, in the region where the solid is less dense than the liquid. A total of 74 simulations have been conducted over the range $\Pstar=16.5$ to $\Pstar=26.0$, $\Tstar=1.42$ to $1.71$. Each simulation is of duration $\tstar=68$ at equilibrium with a system size $N=500$. No anomalies are observed in the density (Fig. \ref{fig:denmax}), heat capacity, bulk modulus or diffusion coefficient. This is to be expected. In contrast to the two-dimensional case in ref \cite{WildingM02}, the melting transition is highly first-order. Nucleation of solid clusters is therefore strongly inhibited by finite-size effects. Accelerated sampling methods may prove useful in this region.
It should be noted that the isostructural transition in the $A=3\epsilon/2$ potential terminates at the ld-fcc spinodal line \emph{below} the glass temperature. This has been located using the method in ref \cite{Wendt78} as $\Tstar_{g}\simeq 0.47$ at $\Pstar=1.5$. No influence on the supercooled liquid is therefore expected. This is confirmed by a sequence of $12$, $N=500$ simulations (also of duration $\tstar=68$) along the $\Tstar=0.44$ and $\Tstar=0.56$ isotherms. No anomalies or evidence of a LLPT are found in the metastable liquid. 

\section{Liquid Structure}

\begin{figure}
\centering
\includegraphics*[scale=0.4]{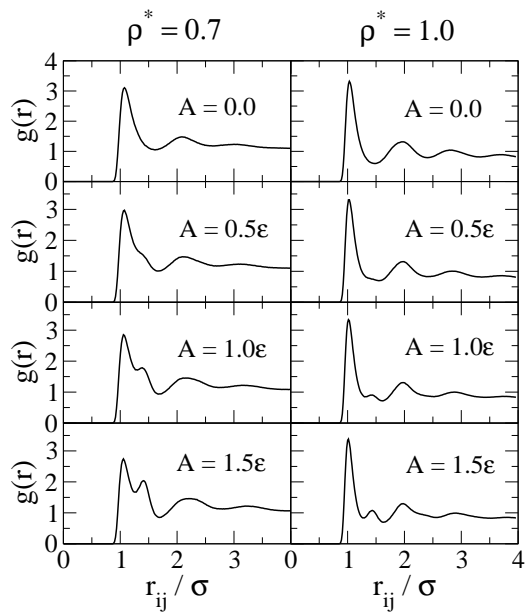}
\caption{Evolution of the pair correlation function $g(r)$ with $A$ at two densities. In each case the temperature is $\Tstar=1.6$.}
\label{fig:rdfs}
\end{figure}

Finally, we briefly examine the structure of the liquid.. The pair correlation function $g(r)$ was computed under a variety of conditions and studied as a function of the parameter $A$. Results at $\Tstar=1.6$ are shown in figure \ref{fig:rdfs}
 for densities $\rho^{*}=0.7$ and $\rho^{*}=1.0$. These were computed from canonical ensemble molecular dynamics simulations using $500$ atoms over a duration of $\tstar=170$.

At the lower density, increasing the depth of the Gaussian well leads to the emergence of a peak not present in the Lennard-Jones ($A=0.0$) case. This corresponds to an coordination shell lying at an intermediate radial distance between the first and second Lennard-Jones shells. The occupation of this shell occurs at the expense of the innermost peak. The effect of increasing the $A$ parameter is hence to shift a significant fraction of first-nearest neighbors into the intermediate coordination shell. The second Lennard-Jones coordination shell is seemingly unaffected.

At the higher density, the first coordination shell is unchanged on increasing $A$. The intermediate peak is less pronounced and emerges at the expense of narrowing the second Lennard-Jones coordination shell. Simulations at intermediate densities indicate that although significant, the change in liquid structure occurs continuously and does not correspond to a phase transition.

\section{Conclusions}

We have mapped the liquid-gas, melting and solid-solid transitions in a sequence of core-softened pair potentials with increasing outer well strengths. The liquid-gas critical temperature has been seen to increase approximately quadratically with increasing $A$. The pressure of the critical point also increases. 

As $A$ is initially increased, the fcc melting temperature is found to reduce, perhaps contrary to the expected behaviour. This decrease in melting temperature can be understood by examining Fig. \ref{fig:SHpots}. For $A$ values in this region, the effect of the Gaussian is to widen the existing Lennard-Jones minimum. This allows larger fluctuations about equilibrium lattice positions for a given temperature. The well known empirical rule of Lindemann \citep{LindMelt} states that melting will occur when the r.m.s. fluctuation is $\sim 15\%$ of the nearest-neighbour distance. This will occur at lower temperatures for wider potential wells.

As the Gaussian outer minimum becomes distinct from the Lennard-Jones minimum, the simple hexagonal structure becomes lower in energy due to a large number of third-nearest neighbours close to $r_{0}$. A transition to fcc under pressure has been observed and is seen to intersect the melting line. The resulting simple hexagonal melting temperature increases with increasing $A$. The pressure of the sh-fcc transition increases with increasing $A$, as expected from the larger enthalpy difference.

We have also seen that the sh structure exhibits a maximum melting temperature, which for larger $A$ values is manifested in the thermodynamically stable regime. The predicted isostructural transition has been observed for the fcc structure only. Energy minima are in fact also observed for two volumes when imposing simple cubic symmetry. The higher energy structure is however mechanically unstable when atoms are permitted to move away from lattice sites. The fcc-fcc transition does not approach the melting line. 

The low pressure phase diagram and sublimation have not been studied here. The phase behaviour is expected to be unchanged from the Lennard-Jones case in this region. The hexagonal close-packed structure has also not been studied in detail. For the range of parameters studied the fcc-fcc transition is not resolvable with the methods used here. The solid should be considered `close-packed' in regions where the fcc structure has been calculated as energetically favourable. We note that the lattice-switch Monte-Carlo method \cite{LSwitchMC1,LSwitchLH} has recently been employed to resolve the fcc-hcp transition in the Lennard-Jones potential \cite{LSwitchLH} and could be used to similar effect in these systems.

The two dominant sources of error in this study are the finite-size error in the solid and the statistical error associated with computing a liquid reference point for thermodynamic integration. The former has been largely corrected for by repeat calculations with larger system sizes. The latter dominates over the liquid finite-size error, but has been controlled to an acceptably low level. All phase boundaries shown can be considered accurate to within temperatures of $\Delta \Tstar=\pm0.05$ and pressures of $\Delta \Pstar=\pm0.1$.

The failure of the isostructural fcc-fcc transition to extend into the supercooled liquid suggests that potentials of this \emph{specific} function form (Lennard-Jones plus Gaussian) are \emph{not} capable of reproducing the unusual phase behaviour of water. Variation of the outer well position and width may yield significantly different phase behaviour to that reported here. The range of parameters over which the energetic ordering of phases is unchanged is currently under investigation. The temperature difference between the ld-fcc spinodal and the glass transition is small. It may therefore be possible to extend the fcc-fcc transition beyond the glass temperature using small changes within this parameter range.

\section{Acknowledgements}

Financial support for this research has been partly provided by an EPSRC studentship. 
Computational facilities were provided under EPSRC grant R47769. The authors would like
to thank Prof. C. Vega for useful discussions.


\end{document}